\documentclass[11pt]{article}
\usepackage[dvips]{graphicx,psfrag}
\usepackage{amssymb}
\usepackage{color}
\input epsf
\definecolor{Blue}{rgb}{0.3,0.3,0.9}
\definecolor{Red}{rgb}{1,0,0}
\definecolor{Green}{rgb}{0,1,0}
\newcommand{\be}{\begin{equation}}
\newcommand{\ee}{\end{equation}}
\newcommand{\bea}{\begin{eqnarray}}
\newcommand{\eea}{\end{eqnarray}}

\begin{document}
\begin{center}
{\Large\bf Quantum gravity in the sky}

\bigskip \bigskip \medskip
{\large Aur\'elien Barrau}
\\[.5cm]
{\it Laboratoire de Physique Subatomique et de Cosmologie, UJF, CNRS/IN2P3, INPG\\
53, av. des Martyrs, 38026 Grenoble cedex, France\\
Institut Universitaire de France\\
email: Aurelien.Barrau@cern.ch}\\

\bigskip \bigskip \medskip
{\large Julien Grain}
\\[.5cm]
{\it Institut d'Astrophysique Spatiale, Universit\'e Paris-Sud 11, CNRS \\ B\^atiments 120-121, 91405
Orsay Cedex, France\\
email: julien.grain@ias.u-psud.fr}\\[1cm]

{\bf Abstract}
\end{center}
Quantum gravity is known to be mostly a kind of metaphysical speculation. 
In this brief essay, we try to argue that, although still extremely difficult to
reach, observational signatures can in fact be expected. The early universe is an
invaluable laboratory to probe "Planck scale physics". With the example of Loop
Quantum Gravity, we detail some expected features.\\

{\it Brief essay written for the "Gravity Research Foundation" and updated as an introduction for a
lecture on the philosophy of gravity.}
\\[1cm]

Building a quantum theory of gravity --that is a quantum model of spacetime itself-- is often
considered as the most outstanding problem of contemporary physics. Why ? Because this is the
unavoidable horizon for unification, explain most scientists. This might not be that clear.
Unquestionably, unification has been an efficient guide for centuries: it worked with Kepler,
with Maxwell, with Glashow, Salam and Weinberg. Yet, is the World more "unified" today than at the
end of the nineteenth century ? Let us consider the macrocosm : stars, planets, comets, dust,
cosmic-rays, pulsars, quasars, white dwarfs, black holes, magnetohydrodynamics turbulence, galaxy
collisions... Is this a "unified" firmament ? Naturally, one should better consider the microcosm.
However, there are today about 120 degrees of freedom in the standard model (not even mentioning
supersymmetry), which is slightly {\it more} than the number of atoms in the old Mendeleev periodic
table. Of course, it might be more relevant to consider interactions instead of matter constituents. 
But, once
again, grand unification is still missing and, to account for the accelerated expansion of the
universe, many of us rely on a quintessence scenario \footnote{This is obviously not the only
possibility. A true cosmological constant, as advocated in \cite{carlo1}, is even possible although
the numerical value doesn't fit quantum mechanical expectations.}. Quintessence means 
{\it quintus essentia}, that is "fifth force". We will not elaborate here on the
string theory landscape \cite{land} which, interestingly, exhibits an 
unprecedented diversity within the realm of a tentative fully unified model\footnote{For a more
philosophical investigation of diversity and unification in physics, one can consider
\cite{aurel}.}. The roads toward unification are probably much more intricate that usually thought: 
they might very well be organized as a kind of {\it rhizome}\footnote{We refer here to the
philosophical concept of the "image of thought", as suggested, within the so-called {\it French
Theory}, in \cite{deleuze}.}.

Does this mean that the idea of quantum gravity itself should be forgotten ? After all, this has been
shown to be such an incredibly difficult theory to establish that the wise attitude could just 
be to withdraw from this apparently never-ending quest. Unfortunately, it is impossible to 
ignore the "gravity-quantum" tension\footnote{A possible way out might still be conceivable in the framework of "emergent" or
"entropic" gravity \cite{ver,jac}.}. The first reason is that the quantum world interacts with
the gravitational field. This, in itself, requires gravity to be understood in a quantum
paradigm, as can be demonstrated by appropriate thought experiments \cite{oriti}. The second reason is that, the other way round, gravity
requires quantum field theories to live in curved spaces. Just because of the equivalence
principle, it is easy to get convinced \cite{birell} that this cannot be
rigorously implemented without quantum general relativity. Basically speaking, the nonlinearity of gravity
frustrates all attempts to ignore quantum gravity: each time a strong gravitational field is involved the coupling
to gravitons should also be strong. The third reason is the existence of singularities: general
relativity predicts, by itself, its own breakdown (as exhibited, {\it e.g.}, by the Penrose-Hawking theorems \cite{singul}). 
This is a truly remarkable feature. Although the first two reasons can, to some extent, be considered
as "heuristic" motivations, the last one does imperatively require a way out. Pure general relativity
ontologically fails. \\

It is sometimes said either that we have too many candidate theories \cite{kiefer} for quantum gravity or that we don't
have a single (convincing) one \cite{rovelli}. Although apparently contradictory, both statements
are in fact correct. This is a quite
specific situation: many different approaches are investigated, all are promising, but none is 
fully consistent. 
At this stage, experiments are obviously missing to eradicate those theories
that are deeply on the wrong track and to improve those that might be correct. Unfortunately, 
quantum gravity is known to be out of reach of any possible experiment\footnote{The ratio of the Planck scale
to the LHC scale is roughly the same than the ratio of the human scale to the distance to the closest
star.}. We shall now underline that this might not be true. \\

\begin{figure}
\begin{center}
\includegraphics[width=110mm]{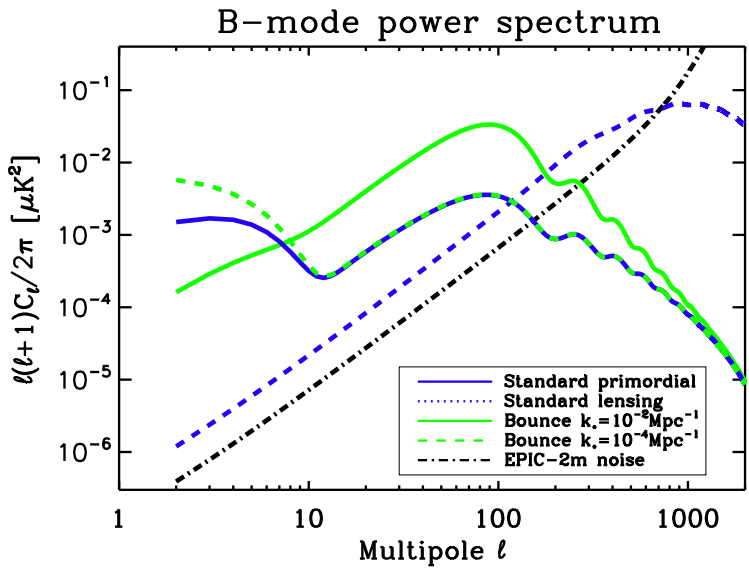}
 \caption{B-mode power spectrum in the "standard model" and with a LQC-induced
 bounce. The lensing background and the sensitivity of an Epic-like experiment are
 also displayed to underline the detectability.}
\label{fig1}
\end{center}
\end{figure}

As previously stated, singularities are the "places" were the departure of quantum gravity from
standard predictions is expected to be the largest (namely infinite). Curing
singular pathologies is indeed the first {\it requisit} for any tentative theory of quantum gravity. 
This qualifies  black holes  and the Big Bang neighborhood as ideal places for quantum gravity 
investigations. From now on, we focus on Loop
Quantum Gravity (LQG) as a prototype model for a background-independent and non-perturbative quantization 
of general relativity \cite{rovelli,lqg}. Clear predictions can be made for the spectrum of black holes 
\cite{bh}, leading to specific signatures in their Hawking evaporation 
products \cite{barrau}. Although tantalizing, this approach\footnote{A very interesting alternative
to probe quantum gravity would be to search for a violation of the Lorentz invariance using photons
emitted, {\it e.g.}, by a gamma-ray burst \cite{liv}. This is extremely motivating but not conclusive
as LQG -among others- do {\it not} predict any clear violation of the Lorentz invariance \cite{lqgliv}.}
is unfortunately not very promising as no
light black hole has been detected so far \cite{pbh} (unless large extra-dimensions are assumed, producing
small black holes is extremely difficult and would typically require a density contrast in the early
universe $10^4$ times larger than measured). We are left with cosmology and the Big Bang singularity resolution.
This is precisely the key prediction of LQG as applied to the universe as a whole -- the so-called Loop 
Quantum Cosmology (LQC)
model \cite{lqc}: the Big Bang is replaced by a Big Bounce.\\

\begin{figure}
\begin{center}
\includegraphics[width=110mm]{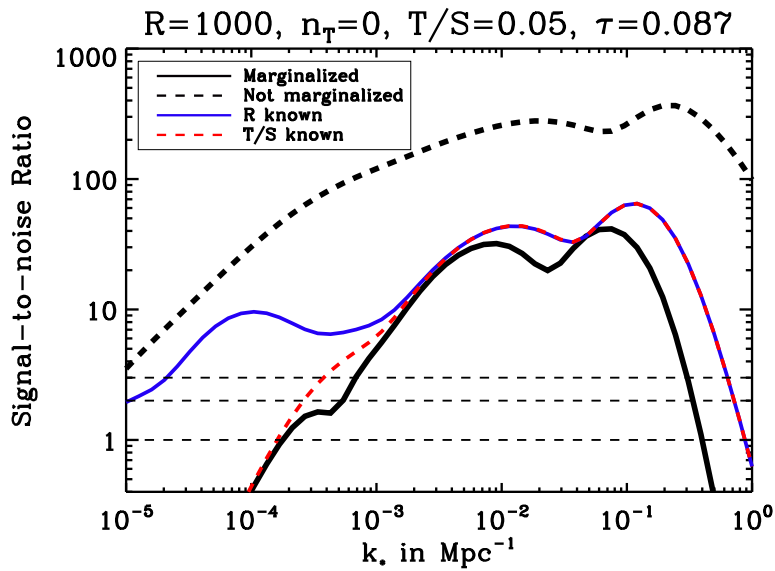}
\caption{Signal-to-noise ratio for the detection of a LQC bounce as a function of the $k_*$ parameter, which determines
the transition between the suppressed spectrum and the standard spectrum, for different marginalization schemes.}
\label{fig2}
\end{center}
\end{figure}

Fortunately, it is now becoming clear that Loop Quantum Cosmology (LQC) provides much more than an elegant
smoothing of the primordial singularity. It somehow {\it predicts} inflation. We will resist the temptation to
compute the accurate probability for inflation to occur in this model \cite{ashtk}, as it highly 
depends on the chosen measure and, {\it a fortiori}, to compute the probability for the Universe to be compatible
with WMAP data \cite{ashtk2}: there are contingent phenomena in the Universe and, as long as data are added or refined, 
the probability will inevitably decrease, even if the theory is correct.
The fact remains that \cite{jakub}, even for the simplest model,
without any fine-tuning (say for a universe filled with a massive scalar
field), the usual "friction" term of the Klein Gordon equation ($\ddot{\phi}+3H\dot{\phi}+m^2\phi=0$) becomes an 
anti-friction one in the contracting phase
of the universe, therefore obliging the field to climb up its potential\footnote{In our opinion, establishing that this
result remains true when anisotropies --that are growing faster than anything else in the contracting phase-- are taken
into account
is the major challenge of this model for the forthcoming years (some preliminary encouraging results are already 
available \cite{we}).} ! Just after
the bounce, the Universe begins to expand, the Hubble parameter becomes positive and therefore acts  as friction:
the field is nearly frozen and  standard "slow-roll" inflation takes place. This is a good point: the canonical 
quantization of general relativity "\`a la loop" makes inflation much more natural than in the standard cosmological 
framework (see \cite{stein} for some "naturality" problems of the inflationary paradigm). This is
however not enough to qualify cosmology as a probe for (loop) quantum gravity. 

\begin{figure}
\begin{center}
\includegraphics[width=110mm]{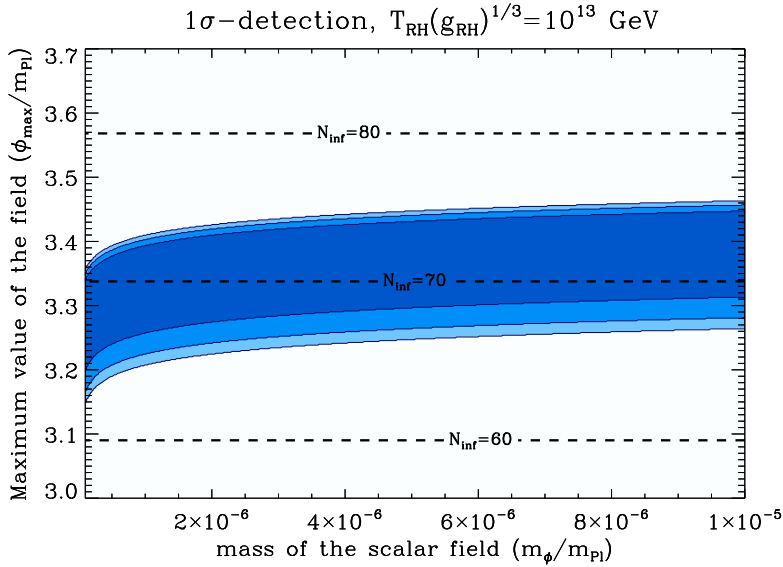}
\caption{Detectability (1, 2 and 3 $\sigma$) as a function of the mass of the field and of the maximum energy reached just
after the bounce.}
\label{fig3}
\end{center}
\end{figure}

The situation dramatically changes when one investigates a bit more into the details the propagation of perturbations
 through
the bounce\footnote{Perturbation theory and averaging 
cosmology are deeply interconnected. The key point is that statistical properties can now be accurately computed.}. 
It is often argued that inflation erases all possible footprints of quantum gravity due to the huge
increase of the scale factor. This is not entirely true. The deep reason for this is the following: in a "Fock space" language,
the occupation numbers {\it after} inflation are  sensitive to the occupation numbers {\it before} inflation, just because
the quantity which is amplified is the number of particles per unit cell of phase space, whose volume ${\rm d}^3x{\rm
d}^3k/(2\pi)^3$ is unaffected by the huge increase of the scale factor (whereas
${\rm d}^3x \propto a^3$).

The basic equation to be solved to investigate gravity waves is the following\footnote{For the sake of simplicity
we have here just considered the so-called "holonomy" correction. The other main LQC term, the inverse triad correction,
does not change the qualitative analysis presented in this note.}:
\begin{equation}
\frac{d^2}{d\eta^2}h^i_a + 2\mathcal{H}\frac{d}{d\eta}h^i_a-\nabla^2h^i_a+
m^2_Qh^i_a = 0,
\end{equation}
where $h^i_a$ are gravitational perturbations, $\eta$ and $\mathcal{H}$ are
the conformal time and  Hubble constant.  The effective mass
term, which encodes LQC corrections, is given by  
\begin{equation}
m^2_Q := 16 \pi G a^2 \frac{\rho}{\rho_c}\left(\frac{2}{3}\rho-V \right),       
\end{equation}
with $\rho_c\approx\rho_{Pl}$, whereas the background follows:
\begin{equation}
H^2 =\frac{\kappa}{3} \rho \left(1-\frac{\rho}{\rho_c}  \right). \label{Friedmann}
\end{equation}

The numerical resolution leads to a power spectrum which is quadratically suppressed below some scale $k_*$, exhibits
a bump of amplitude $R$ around $k_*$, and then follows, in the ultra-violet limit, the (nearly) scale-invariant behavior. This 
can be easily understood: large physical scales ($k<k_*$)  crossed the horizon only once and were frozen in
the Minkowski vacuum ($P(k)\propto k^2$), whereas small scales ($k>k_*$) followed a nearly standard history (they exited 
the horizon during inflation and re-entered later). This deformed primordial power spectrum can be used as an input to estimate 
the resulting B-mode Cosmic Microwave Background (CMB) spectrum (Fig. \ref{fig1}). It exhibits some
deviations with respect to the
usual picture that could be probed by the next-generation cosmology experiments
(either Planck or, in the long run, with a polarization-dedicated experiment). Using
this spectra and taking into account both the instrumental noise and the  astrophysical noise, we have
estimated with a Fisher analysis, as shown in Fig.~\ref{fig2}, the signal-to-noise ratio as a function of the $k_*$ scale.
Furthermore, those parameters can be translated into more fundamental ones, as exhibited on Fig.~\ref{fig3}: the value of the
field at the beginning of inflation (or, alternatively, the fraction of potential energy at the bounce) and its mass.
Obviously, a window is opened for detection (see \cite{lqc_grav} for details on the material used
to build up this picture).\\

Of course, this approach is far from being perfect or fully convincing. The parameter space that can be probed remains quite
limited, the backreaction effects are still neglected, and, most importantly, the temperature CMB spectrum --which is 
already observed-- has not
yet been fully computed (see \cite{lqc_scal} for recent progress). Not to mention that  the algebra of
constraints for tensor modes should be non-trivially modified due to the consistency of Poisson brackets derived for
scalar modes. The signature might even change and become euclidean near the bounce. However, with the example of LQG, it seems
that quantum gravity, which has long been thought to be "untestable", might become an observational science\footnote{This is, of
course, not the only way to test this kind of models thanks to the CMB: non-gaussianities (see, {\it e.g.}, \cite{ng} for a
review) or low-variance circles \cite{circles} are other
possible probes.}. String theory has also interesting predictions for the CMB, {\it e.g.} in the Ekpyrotic scenario \cite{ek}, 
in string gas cosmology \cite{sgc}, or in pre Big Bang cosmology \cite{ppb}, to cite only a few models. 
In the sky, quantum gravity could be confronted with predictions!  Astronomy has already been
extraordinary useful to discover new phenomena. Even with planets alone: while anomalies in the orbit of
Uranus led to the discovery of Neptun, anomalies in the orbit of Mercury required the discovery of
general relativity. Either by adding new constituents or by changing the laws of physics, explaining
data from the sky remains one of the greatest challenges and best insight toward "new physics".

\end{document}